\documentclass[aps, prapplied, twocolumn, showpacs, table]{revtex4-1}
\usepackage[T1]{fontenc}
\usepackage[utf8]{inputenc}
\usepackage{graphicx}
\usepackage{pstricks}
\usepackage{array}
\usepackage{natbib}
\usepackage{amsmath}
\usepackage{pifont}
\usepackage{dsfont}
\usepackage{multirow}
\usepackage{amssymb}
\usepackage{wasysym}
\usepackage[titletoc,toc,title]{appendix}
\usepackage{booktabs}
\usepackage{colortbl}
\usepackage{xcolor}
\usepackage{mathtools}
\usepackage{gensymb}
\usepackage{breqn}
\usepackage{sidecap}
\usepackage{footmisc}

\def\be{\begin{equation}}
\def\ee{\end{equation}}
\newcommand{\eq}[1]{Eq.~(\ref{#1})}

\newcommand{\cnst}{Center for Nanoscale Science and Technology, National
Institute of Standards and Technology, Gaithersburg, MD 20899, USA}
\newcommand{\umd}{Maryland NanoCenter, University of Maryland, College
Park, MD 20742, USA}
\newcommand{\purdue}{School of Electrical \& Computer Engineering, Purdue University, West Lafayette, IN, 47907, USA}

\begin{document}

\title{Sesame: a 2-dimensional solar cell modeling tool}

\begin{abstract}
This work introduces a new software package ``Sesame'' for the numerical computation of classical semiconductor equations.  It supports 1 and 2-dimensional systems and provides tools to easily implement extended defects such as grain boundaries or sample surfaces.  Sesame has been designed to facilitate fast exploration of the system parameter space and to visualize local charge transport properties. Sesame is distributed as a Python package or as a standalone GUI application, and is available at \url{https://pages.nist.gov/sesame/}.
\end{abstract}

\author{Benoit Gaury}
\affiliation{\cnst}
\affiliation{\umd}
\author{Yubo Sun}
\affiliation{\purdue}
\author{Peter Bermel}
\affiliation{\purdue}
\author{Paul M. Haney}
\affiliation{\cnst}

\date{\today}

\maketitle

\section{Introduction}

Numerical simulations are an essential aspect of photovoltaic research and design.  A number of free software packages have been developed and extensively used for solar cell modeling in 1-dimension, including AMPS [1], PC-1D \cite{basore1996pc1d}, SCAPS \cite{burgelman2000modelling}, and wxAMPS \cite{liu2012new}.  Freely available 2-dimensional simulation tools are less common \cite{gray1991adept,altermatt2011models,basore2011pc2d}, but are necessary for describing systems with lateral inhomogeneity.  A common class of such systems are polycrystalline thin film photovoltaics, such as CdTe \cite{major2016grain}, CIGS \cite{yan2006grain}, and hybrid perovskites \cite{yun2015benefit}.  In these materials grain boundaries break the lateral symmetry of the $p$-$n$ junction, leading to complex system geometries.  Lateral inhomogeneity is also often encountered in nanoscale or mesoscopic measurements.  The resolution of these measurement is typically achieved using an excitation source or measurement probe with nanoscale spatial extent.  Examples include electron beam induced current (EBIC) or scanning Kelvin probe microscopy, which are also often surface sensitive.  An appropriate model for these measurements is therefore (at least) 2-dimensional and includes localized excitation/detection sources and relevant boundary conditions.

\begin{figure}
\includegraphics[width=0.3\textwidth]{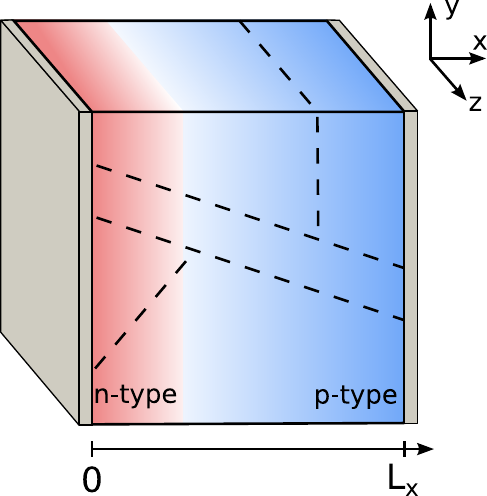}
\caption{\label{geometry} Coordinate system of Sesame: rectilinear geometry and contacts located at $x=0$ and $x=L$.  $pn$ junction doping is shown as an example.}
\end{figure}

There are numerous examples of the use of 2-dimensional solar cell modeling in the literature.  For instance, the impact of grain boundaries in polycrystalline cells has been previously studied numerically \cite{edmiston1996improved,gloeckler2005grain,rau2009grain} and analytically \cite{gaury2016charged,gaury2017charged}.  Simulations have been used for interpreting experiments with localized excitations such as EBIC \cite{haney2016depletion,jin2017impact}, cathodololuminescence \cite{kanevce2015quantitative,mendis2016role}, and two-photon photoluminescence \cite{kanevce2014role}.  Although these works are instructive, nonlinearities in the system response prevent a simple extrapolation of previous results to all possible system configurations of interest.  Indeed there remain a number of unresolved questions of fundamental interest in polycrystalline photovoltaics, questions as basic as whether grain boundaries are harmful or beneficial to cell performance \cite{major2016grain,kumar2014physics}.  It is therefore desirable for researchers to have widespread access to 2-d simulation software.

In this work we introduce Sesame, a Python package developed by the authors (B. G. and P. M. H.) which solves the drift-diffusion-Poisson equations in 1 and 2 dimensions.  Sesame is open source and distributed under the BSD license.  Sesame is designed to easily construct systems with planar defects, such as grain boundaries or sample surfaces, which may contain both discrete or a continuum of gap state defects.  While full-featured commercial packages allow simulations of complex device configurations together with multiple physical effects, the needs of research sometimes require access to the source code and licensing that enables usage on computing clusters.  The program and its code are publicly available at \url{https://pages.nist.gov/sesame/}.

The paper is organized as follows: In Sec. II we present a brief overview of the model and geometry.  In Sec. III we compare the output of Sesame to established  semiconductor modeling software, including SCAPS \cite{burgelman2000modelling}, Sentaurus \cite{guide2013synopsys}, and COMSOL Semiconductor Module \cite{multiphysics2017v} \cite{disclaimer}. In Sec. III we also present a hands-on tutorial script for solving a 2-dimensional system with a grain boundary, and briefly describe the functionality of the GUI.  The mathematics underlying the model and technical details of the numerical implementation can be found in the Appendix.

\section{Overview of the physical model}

The system geometry consists of a semiconductor device connected to contacts at $x=0$ and $x=L$ (see Fig.~\ref{geometry}).  Sesame describes the steady state behavior of this system, which is governed by the drift-diffusion-Poisson equations:

\begin{align}
   \label{ddp1}
   \vec{\nabla}\cdot \vec{J}_n &= -q(G-R)\\
   \label{ddp2}
   \vec{\nabla}\cdot \vec{J}_p &= q(G-R)\\
   \label{ddp3}
   \vec{\nabla}\cdot\left(\epsilon\vec{\nabla}\phi\right) &= -\rho/\epsilon_0
\end{align}
with the currents
\begin{align}
   \label{currents1}
   \vec{J}_n &= -q\mu_n n \vec{\nabla} \phi + qD_n \vec{\nabla}n \\
   \label{currents2}
   \vec{J}_p &= -q\mu_p p \vec{\nabla} \phi - qD_p \vec{\nabla}p~,
\end{align}
where $n$ and $p$ are the respective electron and hole number densities, and $\phi$ is the electrostatic potential. $\vec{J}_{n(p)}$ is the charge current density of electrons (holes). Here, $q$ is the absolute value of the electron charge. $\rho$ is the local charge density, $\epsilon$ is the dielectric
constant of the material, and $\epsilon_0$ is the permittivity of free space. $\mu_{n,p}$ is the electron/hole mobility, and is assumed to satisfy the Einstein relation: $D_{n,p} = k_BT\mu_{n,p}/q$.  $G$ is the electron/hole pair generation rate density and $R$ is the recombination rate density.

Sesame includes Schockley-Read-Hall, radiative, and Auger recombination mechanisms. Sesame is currently limited to describing non-degenerate semiconductors with Boltzmann statistics, and does not include thermionic emission and quantum tunneling at interfaces.  These can be important contributions to the transport in heterojunctions \cite{horio1990numerical}, so care should be exercised when using Sesame to simulate such systems.  Sesame includes Ohmic and Schottky contact boundary conditions, and periodic or hardwall (infinite potential) transverse boundary conditions. Sesame uses finite differences to solve Eqs. (\ref{ddp1}-\ref{ddp3}), and the standard Scharfetter-Gummel scheme for discretizing the current \cite{Gummel}.  Details of the implementation can be found in the Appendix.

\section{Benchmarks and Examples}

\subsection{Benchmarks}

\begin{figure}
\includegraphics[width=0.475\textwidth]{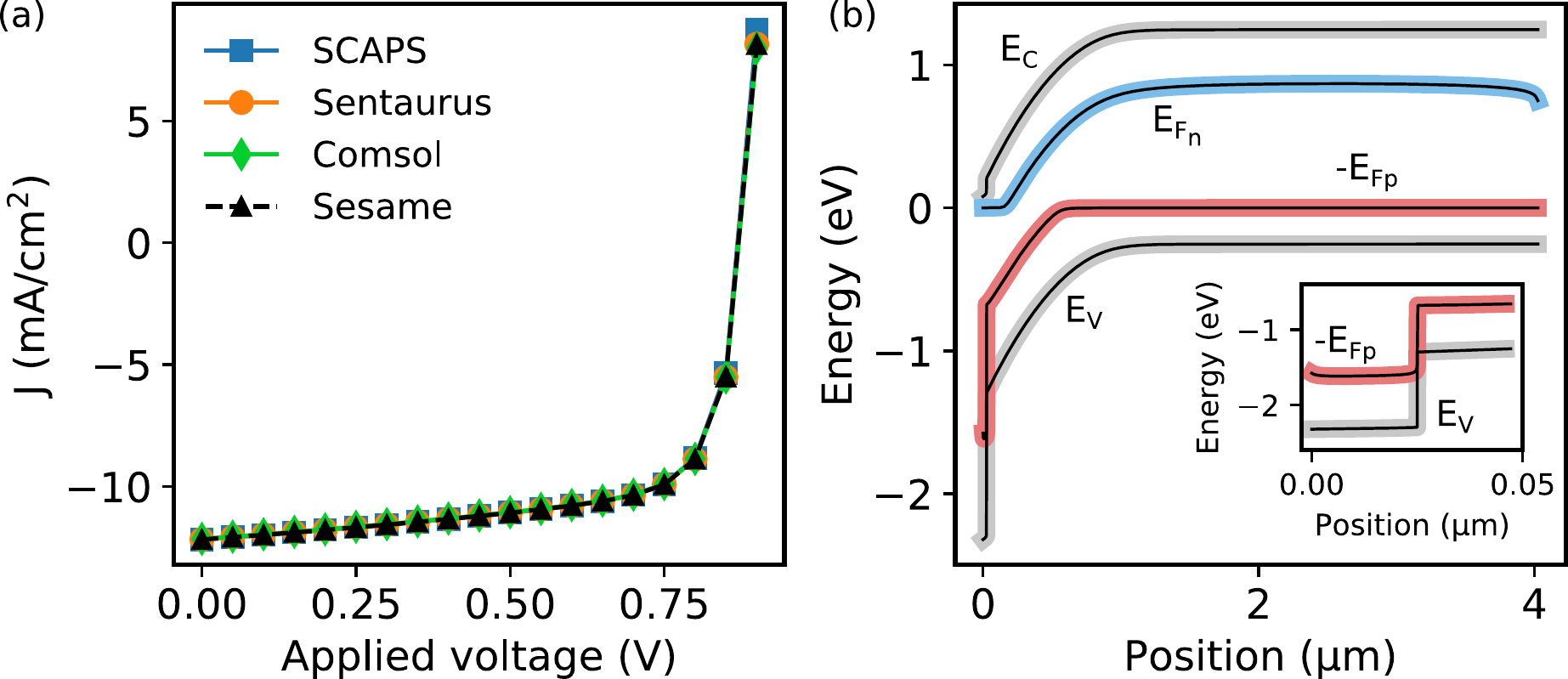}
\caption{\label{fig:energy}Comparison between Sesame and SCAPS, Sentaurus, and COMSOL for a CdS-CdTe heterojunction.  (a) Illuminated $J$-$V$ curve. (b) Band diagram under short circuit conditions.  Inset shows the valence band and hole quasi-Fermi level near the CdS layer.  Black thin lines are the Sentaurus results, and thick colored lines are Sesame results. } \label{fig:1d}
\end{figure}

We first verify the consistency between Sesame and other software packages.  We have compared the output of Sesame with the well-established software packages Sentaurus, COMSOL, and SCAPS for many systems, and present two illustrative examples here.  We first consider a 1-d heterojunction consisting of a thin $n^+$-doped layer of CdS and a $p$-type CdTe.  The material parameters are shown in Table \ref{table:params1}.  Fig \ref{fig:1d}(a) shows the computed $J$-$V$ curve under uniform illumination of $G=3.3\times10^{20}~{\rm cm^{-3}~s^{-1}}$.  We find close agreement between Sesame, Sentaurus, and COMSOL.  To quantify the comparison, we define the relative difference between two computed currents $J_1$ and $J_2$ as $\left|J_1-J_2\right|/\langle J_1+J_2\rangle$, where $\langle\rangle$ denotes the average.  The maximum relative difference between Sesame and Sentaurus is $0.2~\%$, and between Sesame and COMSOL it is $2~\%$.  We observe a more substantial difference between Sesame and SCAPS, with a maximum value of $7~\%$.  In all cases, the maximum discrepancy occurs near $V_{\rm oc}$, where the current is minimized so that {\it relative} differences are maximized.  We attribute the larger difference between Sesame and SCAPS to the different interface recombination model used in SCAPS, in which the system variables are multi-valued at the interface and allow for recombination between layers \cite{burgelman2000modelling}.

\begin{table}
\begin{tabular}{|l|c|c|}
  \hline
  Param. & CdS & CdTe  \\ \hline
  $L~ [{\rm nm}]$ & $25$ & $4000$ \\
  $\epsilon$ & 10 & 9.4\\
  $\tau_n~[{\rm ns}]$ & $ 10$ & $5$ \\
  $\tau_p~[{\rm ns}]$ & $ 10^{-4}$ & $5$ \\
  $N_C~[{\rm cm^{-3}}]$ & $2.2\times10^{18}$ &  $8\times10^{17}$\\
  $N_V~[{\rm cm^{-3}}]$ & $1.8\times10^{19}$ &  $1.8\times10^{19}$\\
  $E_g~[{\rm eV}]$ & $2.4$  &  $1.5$\\
  $\chi~[{\rm eV}]$ & $4.0$  &  $3.9$\\
  $\mu_n~[{\rm cm^2/\left(V\cdot s\right)}]$ & $100$ & $320$\\
  $\mu_p~[{\rm cm^2/\left(V\cdot s\right)}]$ & $25$ & $40$\\
  doping$~[{\rm cm^{-3}}]$ & $10^{17}$ (D) &  $10^{15}$ (A)\\ \hline
\end{tabular}
\caption{List of parameters used for the 1-d heterojunction calculation.  The label (D) and (A) for the doping value indicate donor and acceptor, respectively. \label{table:params1} }
\end{table}

\begin{figure}
\includegraphics[width=0.475\textwidth]{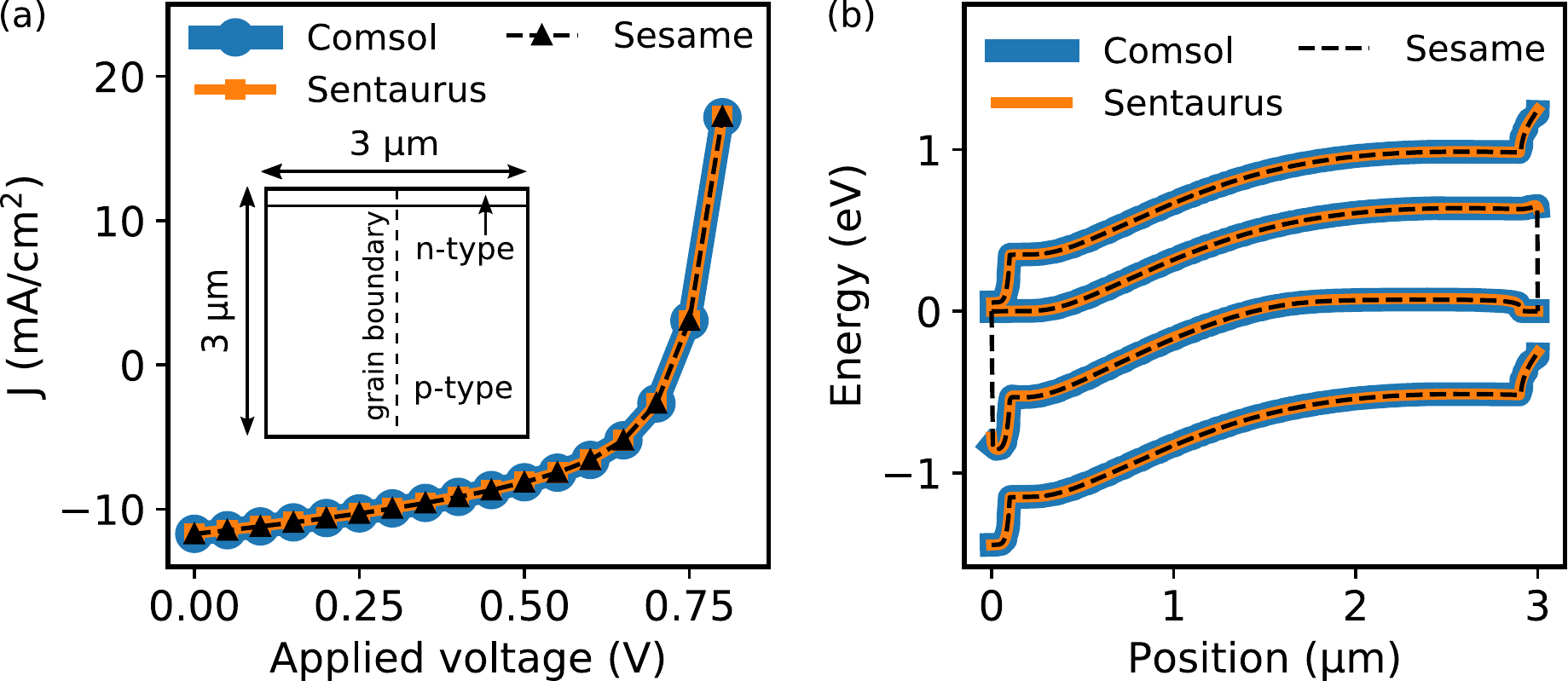}
\caption{\label{fig:energy}Comparison between Sesame, COMSOL and Sentaurus for a 2-dimensional system. (a) Illuminated JV curve. Inset: schematic of the system, an n-p junction with a columnar grain boundary. (b) Band diagram along the grain boundary core under short-circuit conditions.   } \label{fig:2d}
\end{figure}

We next consider a 2-d homojunction with a single columnar grain boundary (see inset of Fig. \ref{fig:2d}(a) for system geometry).  We use the same bulk parameters as given for CdTe in Table \ref{table:params1} for both $n$ and $p$ layers, except with $\tau_e=\tau_h=10~{\rm ns}$.  The thickness of the $n^+$ layer is taken to be $100~{\rm nm}$.  The grain boundary contains a donor and an acceptor defect, both positioned at $0.4~{\rm eV}$ above midgap, with defect density $\rho_{GB}=10^{14}~{\rm cm^{-2}}$ and equal hole and electron capture cross section $\sigma_{\rm GB}=10^{-14}~{\rm cm^{2}}$.  The grain boundary is positioned in the middle of the system, and terminates at a distance of $100~{\rm nm}$ from either contact.  For this simulation we again use a uniform generation rate $G=3.3\times 10^{21}~{\rm cm^{-3}~s^{-1}}$.  Fig. \ref{fig:2d}(a) shows the illuminated $J$-$V$ curve obtained with Sesame, COMSOL, and Sentaurus (SCAPS is not included, as it does not support 2-d geometries). We again find good agreement between Sesame and the other software packages: the largest relative difference between Sesame and Sentaurus is 0.5~\%, and between Sesame and COMSOL it is 0.7~\%.  Fig. \ref{fig:2d}(b) shows the good agreement obtained for the band diagram along the grain boundary core under short-circuit conditions for the three software packages.

\subsection{Scripting example}

Sesame is run either through a self-contained GUI, or as a python package which is called in scripts.  Running sesame with scripts is particularly convenient for running large-scale batch simulations on a computing cluster.  Scripting also provides more flexibility in system definition ({\it e.g.} continuous grading of electronic parameters and doping).  In the distribution, we provide several example scripts which describe standard PV simulations (e.g. $J$-$V$, IQE calculations), along with in-depth tutorials in the documentation.  Here we give a description of the script used to generate the data of Fig. \ref{fig:2d}.

We first import the numpy and sesame packages:\\
\\
{\fontfamily{qcr}\selectfont
\small{
\noindent import sesame\\
import numpy as np\\
}
}
\\
\noindent Next we define the grids for $x$ and $y$.  We use uniform grids for this example, but generally non-uniform grids are necessary to optimize the simulation accuracy and speed (non-uniform grids are used in the simulation of Fig. \ref{fig:2d}).  (Note: Sesame assumes all lengths are given in units of cm.)
\\

{\fontfamily{qcr}\selectfont
\small{
\noindent x = np.linspace(0,3e-4,100)  \\
y = np.linspace(0,3e-4,100)
}
}
\bigskip
\\
\noindent
We create the system with the {\fontfamily{qcr}\selectfont\small{Builder}} function.  The input to {\fontfamily{qcr}\selectfont\small{Builder}} are the $x$ and $y$ grids.  The output is an object {\fontfamily{qcr}\selectfont\small{sys}} which contains all the information needed to describe the simulation.
\\

{\fontfamily{qcr}\selectfont
\small{
\noindent sys = sesame.Builder(x, y)
}
}
\bigskip
\\
\noindent Additional simulation settings are set by calling various methods of {\fontfamily{qcr}\selectfont\small{sys}}, as we show below.

Next we define the material properties with a python dictionary object (called {\fontfamily{qcr}\selectfont\small{mat}} in this example).  The dictionary key names correspond to standard definitions. (Note: Sesame assumes times are given in units of ${\rm s}$, energies in units of ${\rm eV}$, densities in units of ${\rm cm^{-3}}$, mobility in units of ${\rm cm/V\cdot s}$)).
\\

{\fontfamily{qcr}\selectfont
\small{
\noindent mat = \{'Nc':8e17, 'Nv':1.8e19, 'Eg':1.5, \\
\indent 'affinity':4.1, 'epsilon':9.4, 'Et':0,\\
\indent 'mu_e':320, 'mu_h':40, 'tau_e':1e-8, \\
\indent 'tau_h':1e-8\}
}
}

\bigskip
\noindent The dictionary key {\fontfamily{qcr}\selectfont\small{Et}} represents the energetic position of bulk recombination centers, as measured from the intrinsic energy level, and {\fontfamily{qcr}\selectfont\small{tau_e}}/{\fontfamily{qcr}\selectfont\small{tau_h}} are the electron/hole lifetimes.  The dependence of the Schockley-Read-Hall recombination on these parameters can be found in the Appendix.  The material is added to the system using the {\fontfamily{qcr}\selectfont\small{add_material}} function, which takes the {\fontfamily{qcr}\selectfont\small{mat}} dictionary as input.  Note that {\fontfamily{qcr}\selectfont\small{add_material}} is a method of the {\fontfamily{qcr}\selectfont\small{sys}} object, and is called with the command:
\\

{\fontfamily{qcr}\selectfont
\small{
\noindent sys.add_material(mat)
}
}

\bigskip
\noindent To build a $p$-$n$ junction we add a position-dependent doping profile to the system.  We must define functions which describe the different doping regions; for this example, these functions are called {\fontfamily{qcr}\selectfont\small{n_region}} and {\fontfamily{qcr}\selectfont\small{p_region}}.  They return True when the input variable {\fontfamily{qcr}\selectfont\small{position}} belongs to the region.  For this example the two regions are delimited at the {\fontfamily{qcr}\selectfont\small{junction}} coordinate which corresponds to $x=10^{-5}~{\rm cm}$.
\bigskip
\\
{\fontfamily{qcr}\selectfont
\small{
\noindent junction = 1e-5    \\
\\
def n_region(position):\\
\indent    x, y = position\\
\indent    return x < junction\\
\\
def p_region(position):\\
    \indent x, y = position\\
    \indent return x >= junction\\
}
}

\noindent Having defined the different doping regions, we add the donors and acceptors with the {\fontfamily{qcr}\selectfont\small{sys}} methods {\fontfamily{qcr}\selectfont\small{add_donor}} and {\fontfamily{qcr}\selectfont\small{add_acceptor}}.  The input for these methods are the doping magnitude and doping region functions we just defined.  Sesame currently assumes that all bulk dopants are fully ionized. (Note: Sesame assumes the units of density is ${\rm cm^{-3}}$):
\bigskip
\\
{\fontfamily{qcr}\selectfont
\small{
donorDensity = 1e17 \quad\quad\quad   \\
sys.add_donor(donorDensity, n_region)  \\
acceporDensity = 1e15\\
sys.add_acceptor(acceptorDensity, p_region)\\
}
}

\noindent Next we specify the contact boundary conditions.  For this example, we specify Ohmic contacts with the function {\fontfamily{qcr}\selectfont\small{contact_type}}.  Note the order of input arguments is left contact ($x=0$) type first, right contact ($x=L$) type second:
\bigskip
\\
{\fontfamily{qcr}\selectfont
\small{
sys.contact_type('Ohmic','Ohmic')\\
}
}

\noindent We next specify the value of recombination velocity for electrons and holes at both contacts (Note: Sesame assumes the units of velocity are ${\rm cm/s}$).  For this example, both contacts only collect majority carriers.  This is accomplished with the function {\fontfamily{qcr}\selectfont\small{contact_S}}:
\bigskip
\\
{\fontfamily{qcr}\selectfont
\small{
Sn_L, Sp_L, Sn_R, Sp_R = 1e7, 0, 0, 1e7\\
sys.contact_S(Sn_L, Sp_L, Sn_R, Sp_R)\\
}
}

\noindent Next we add a grain boundary.  We must specify the grain boundary defect energy level {\fontfamily{qcr}\selectfont\small{EGB}} (note the defect energy level is measured from the intrinsic energy level), the electron and hole capture cross sections {\fontfamily{qcr}\selectfont\small{sigmaeGB}} and {\fontfamily{qcr}\selectfont\small{sigmahGB}}, the defect density {\fontfamily{qcr}\selectfont\small{rhoGB}}, and the endpoints of the line defining the grain boundary {\fontfamily{qcr}\selectfont\small{p1, p2}}.  These are input arguments to the function {\fontfamily{qcr}\selectfont\small{add_line_defects}} which creates a grain boundary.  We also specify the charge transition states of the defect with the function input {\fontfamily{qcr}\selectfont\small{transition}}.  In this case the specified charge states are (+1,-1), corresponding to having a donor and acceptor at the same energy level.
\bigskip
\\
{\fontfamily{qcr}\selectfont
\small{
\noindent EGB = 0.4         \\
sigmaeGB = 1e-15       \\
sigmahGB = 1e-15       \\
rhoGB = 1e14  \\
p1 = (.1e-4, 1.5e-4)   \\
p2 = (2.9e-4, 1.5e-4)  \\
sys.add_line_defects([p1, p2], rhoGB, \\
\indent sigmaeGB, sigmahGB, EGB, transition=(+1,-1)) \\
}
}

\noindent We add illumination by defining a function {\fontfamily{qcr}\selectfont\small{illumination}} which returns the position-dependent intensity as a function of the input coordinate {\fontfamily{qcr}\selectfont\small{x,y}}
\bigskip
\\
{\fontfamily{qcr}\selectfont
\small{
def illumination(x,y):\\
\indent return 2.3e21 * np.exp(-2.3e4 * x)\\
sys.generation(illumination)\\
}
}

\noindent With the system now fully defined, we specify the list of applied voltages used to compute the current-voltage relation with the {\fontfamily{qcr}\selectfont\small{IVcurve}} function:
\bigskip
\\
{\fontfamily{qcr}\selectfont
\small{
voltages = np.linspace(0,1,11)\\
jset = sesame.IVcurve(sys, voltages, \\
\indent solution, 'GB_JV')\\
}
}

\noindent The function {\fontfamily{qcr}\selectfont\small{IVcurve}} returns an array {\fontfamily{qcr}\selectfont\small{jset}} containing the computed current density for each applied voltage.  The {\fontfamily{qcr}\selectfont\small{IVcurve}} function also saves output files with seedname ``GB_JV'' concatenated with a suffix labeling the applied voltage index.  These output files contain objects describing the simulation settings and the solution arrays.  By default these files are compressed data files containing python Pickle objects (.gzip files).  There is also an option to output the data in Matlab format (.mat files).  Sesame includes an {\fontfamily{qcr}\selectfont\small{Analyzer}} object which contains several functions for computing quantities of interest from the solution, such as current densities, total recombination, carrier densities, and others.  We refer the reader to the online documentation for a detailed list of all these functions.

\begin{figure}
\includegraphics[width=0.3\textwidth]{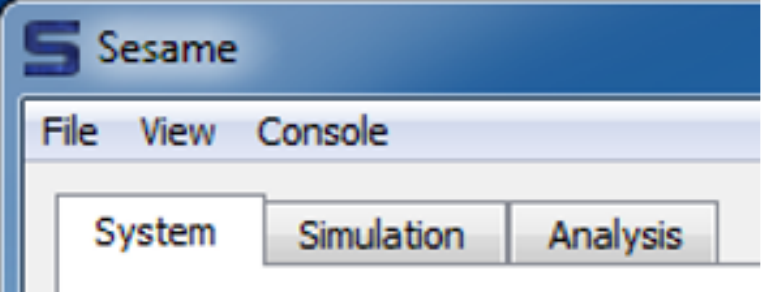}
\caption{Menu and 3 main tabs of the Sesame GUI.  See text for a description of the functionality of each tab.}\label{fig:menu}
\end{figure}

\subsection{GUI}

Use of the standalone GUI as an alternative to scripting can be more convenient for small-scale calculations, or for those without access to a python distribution.  Simulation settings can be saved and loaded, and the GUI also provides an interactive python prompt.  The GUI is divided into three tabs, as shown in Fig. \ref{fig:menu}:\\

\begin{figure}
\includegraphics[width=0.5\textwidth]{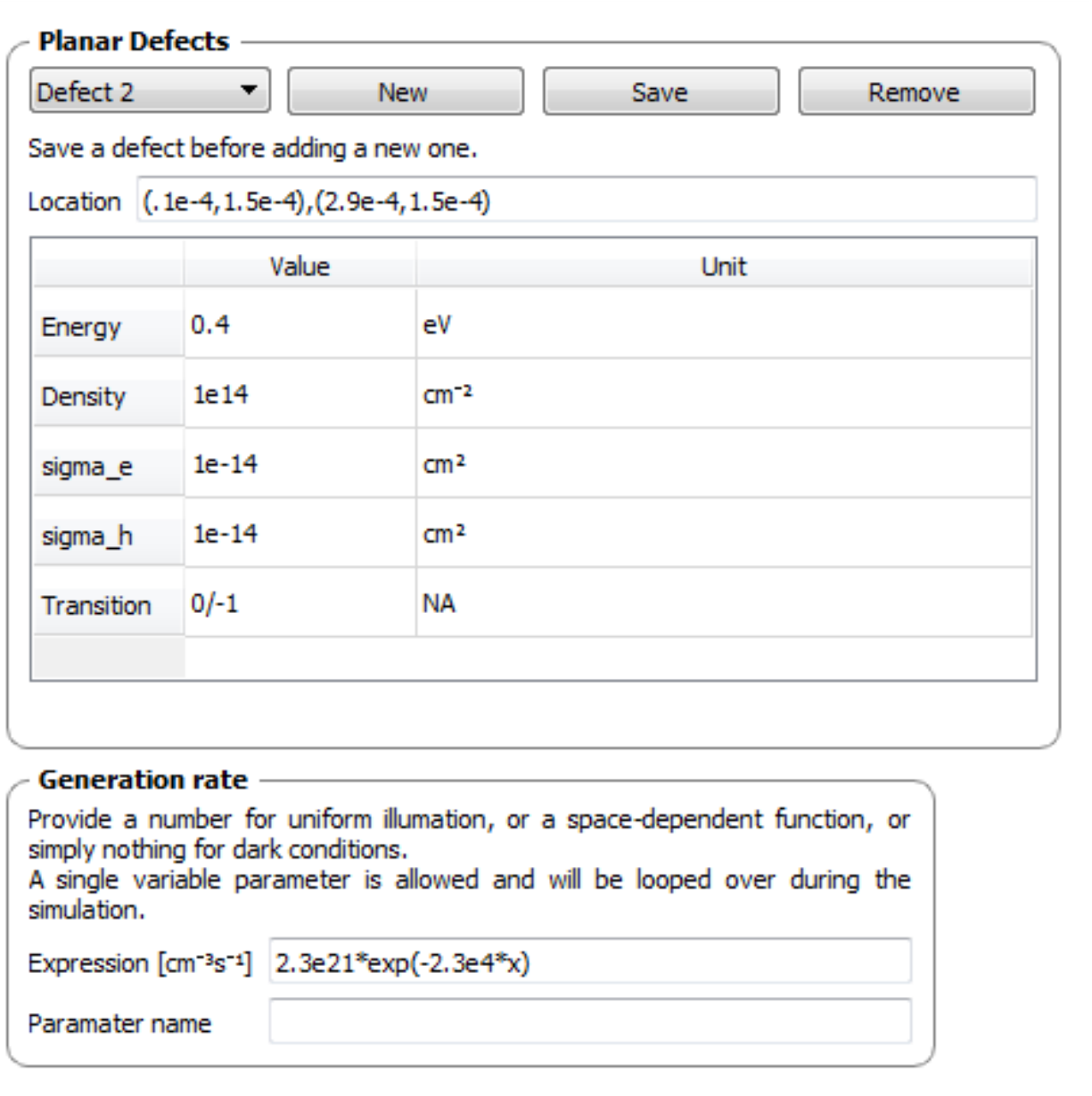}
\caption{\label{fig:sim}Panels from System tab of the GUI.  In the planar defects panel, an arbitrary number of planar defects can be defined by specifying the endpoints of the boundary (a 1-d line for a 2-dimensional simulation), the defect energy, density, electron and hole capture cross sections, and charge states.  The Generation rate panel allows for one user-defined parameter to be varied. }
\end{figure}

\begin{figure}
\includegraphics[width=0.3\textwidth]{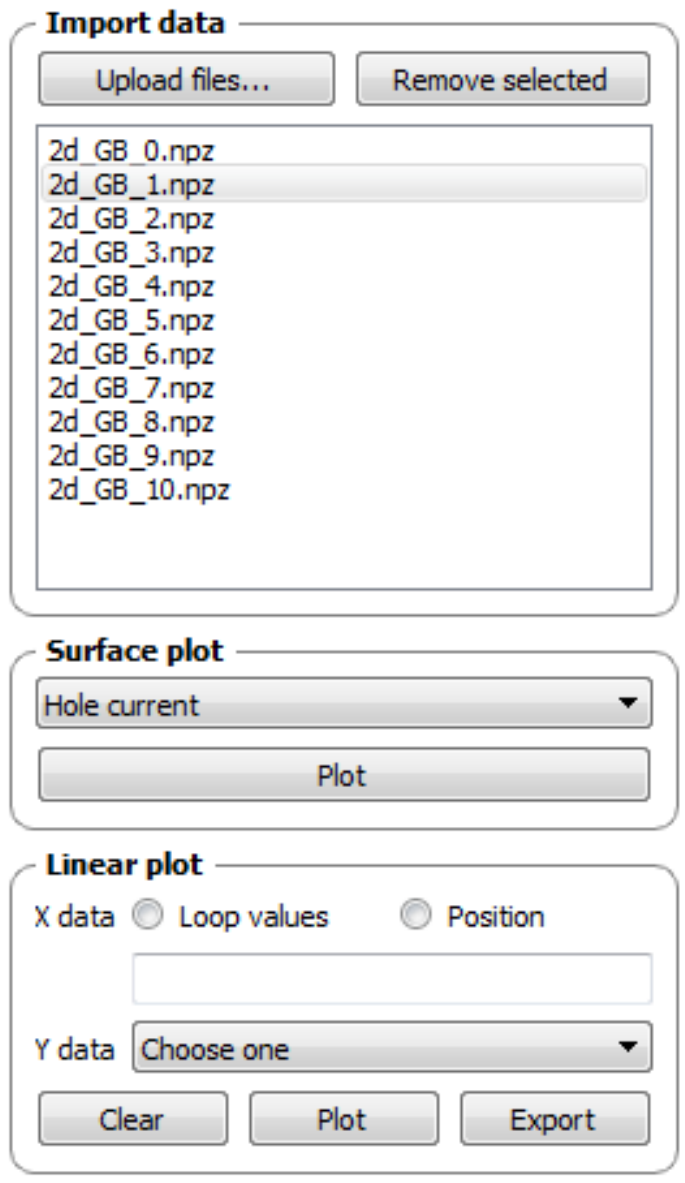}
\caption{\label{fig:import}Panel from Analysis tab of the GUI.  Data files can be selected for analysis, and surface plots of system variables and observables can be generated for 2-dimensional systems.  A linear plot with two modes is available.  In ``loop values'' mode, a scalar (such as total current) is plotted versus the looped parameters.  In the ``position'' mode, a system variable or observable from a single solution is plotted versus spatial coordinate. }
\end{figure}

\begin{figure}
\includegraphics[width=0.4\textwidth]{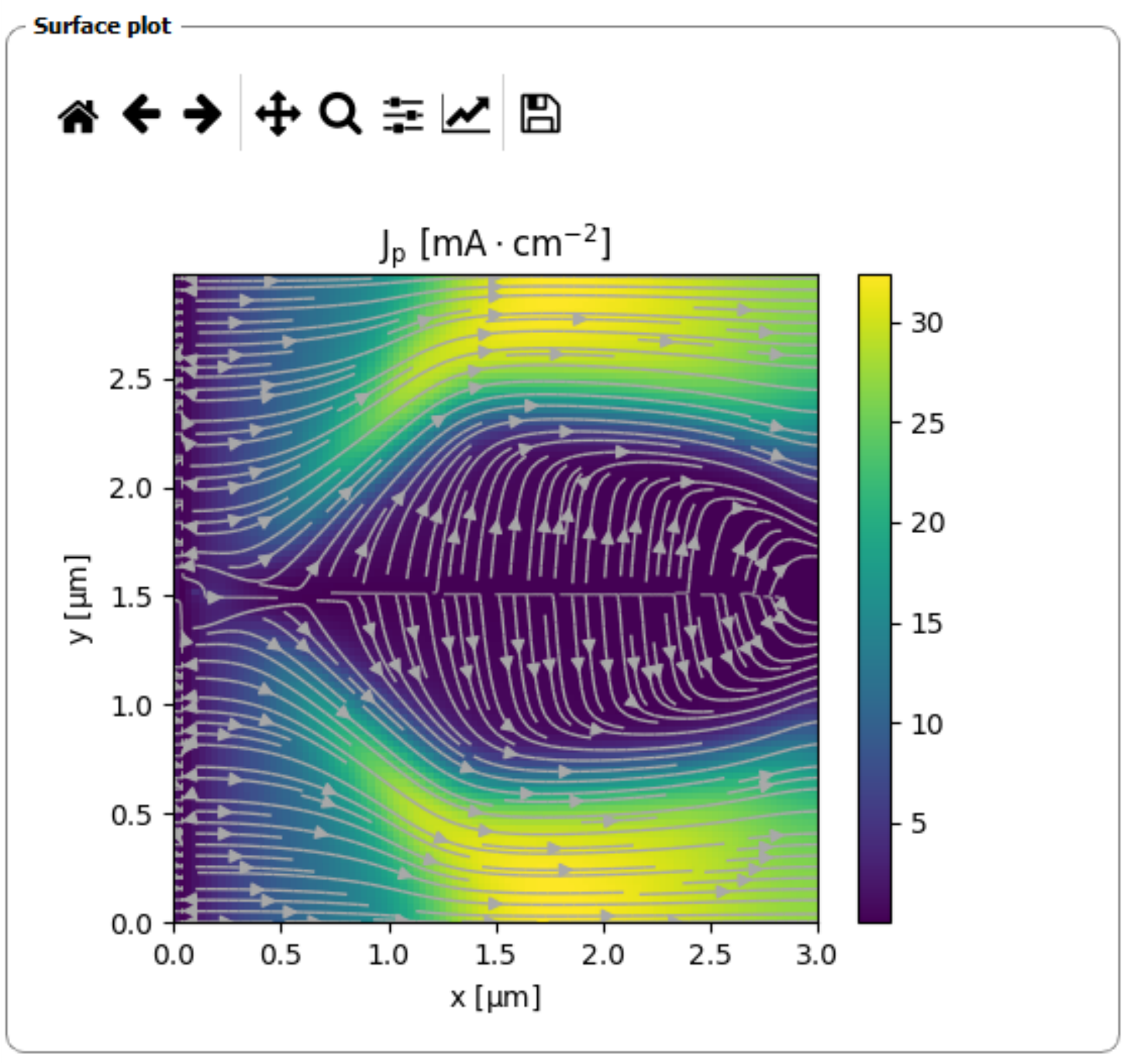}
\caption{\label{fig:current} The surface plot panel of the Analysis tab of the GUI.  This shows the hole currents flowing in a homojunction with a single grain boundary.   }
\end{figure}

1. The System tab contains fields to define the system geometry and material parameters (see Fig. \ref{fig:sim}).\\

2. The Simulation tab lets the user specify which parameter is varied: either the voltage is swept, or a user-defined variable related to the generation rate density is swept.  The boundary conditions and output file information is also set here, and the simulation is launched from this tab.  The program output is provided so that the user can follow the progress of the calculations.\\

3. The Analysis tab enables the user to plot the output of the simulation, and to save and export plotted data (see Figs. \ref{fig:import} and \ref{fig:current}).\\

Sesame is distributed with a number of sample input files for setting up standard PV simulations in the GUI.  More detailed documentation for the GUI is included in the distribution.

\section{Conclusion}

Modeling tools are essential for describing and understanding polycrystalline materials and nanoscale measurements.  System behavior for complex, 2-dimensional geometries can be drastically different than the textbook 1-dimensional $p$-$n$ junction model.  Numerical simulations provide the capability to explore and develop intuition about this rather unchartered territory.  Our aim in releasing Sesame is to provide the research community with a free, easy-to-use resource which will enable broader use of simulation in complex photovoltaic systems.  There are opportunities for additional functionalities (e.g. time-dependence, small-signal analysis, more advanced interface transport models) and further optimizations (e.g. use of Cython) of the code.  Our intent in releasing the fully documented source code is to provide users the option to make these and other additions as their research needs require.  An additional feature not discussed here is 3-dimensional modeling, which is included in the distribution as an untested feature which will be investigated further in future work.

\begin{acknowledgements}
B.~G. acknowledges support under the Cooperative Research
Agreement between the University of Maryland and the National Institute of
Standards and Technology Center for Nanoscale Science and Technology, Award
70NANB14H209, through the University of Maryland.  Y.~S. and P.~B. acknowledge Support provided by the Department of Energy, under DOE Cooperative Agreement No. DE-EE0004946 (PVMI Bay Area PV Consortium), and the National Science Foundation Award EEC 1454315 – CAREER: Thermophotonics for Efficient Harvesting of Waste Heat as Electricity.  We thank Marc Burgelman for helpful correspondence, and thank Mike Scarpulla and Heayoung Yoon for helpful feedback on the software design.
\end{acknowledgements}

\bibliography{ref}

\begin{thebibliography}{32}
\bibitem [{dis()}]{disclaimer}%
  \BibitemOpen
  \href@noop {} {}\bibinfo {note} {The full description of the procedures used
  in this paper requires the identification of certain commercial products. The
  inclusion of such information should in no way be construed as indicating
  that such products are endorsed by NIST or are recommended by NIST or that
  they are necessarily the best software for the purposes
  described.}\BibitemShut {Stop}%
\end{thebibliography}

\begin{appendices}
\section{Model Details}
\label{sec:Model Details}
\subsection{Mathematical Description}
In this section we provide a full description of the equations solved by Sesame.  These are fairly standard and can be found in textbooks \cite{pierret1996semiconductor,fonash1981,selberherr2012analysis,vasileska2017computational}, but we include them here for the sake of completeness and to specify notation and conventions used in the code.  We first write densities in terms of quasi-Fermi levels, denoted by $E_{F_n}$ and $E_{F_p}$ for electrons and holes, respectively.  Since we assume Boltzmann statistics ({\it i.e.} a non-degenerate semiconductor), the carrier densities are related to quasi-Fermi levels by:
\begin{align}
    \label{n}
   n &= N_C \exp\left(\frac{E_{F_n} + \chi + q\phi}{k_BT}\right)\\
   p &= N_V \exp\left(\frac{-E_{F_p} - \chi - E_g - q\phi}{k_BT}\right).
   \label{p}
\end{align}
where $E_g$ is the material band gap, $\chi$ is the electron affinity, and $N_{C,V}$ are the
conduction, valence band effective density of states, respectively.  All quantities except temperature can vary with
position.

The electron and hole current can be expressed in terms of the spatial
gradient of the quasi-Fermi levels \cite{van1980shockley}:
\begin{align}
   \vec{J}_n &= q\mu_n n \vec{\nabla} E_{F_n}\\
   \vec{J}_p &= q\mu_p p \vec{\nabla} E_{F_p}.
\end{align}

\subsubsection{Recombination}
Sesame includes Shockley-Read-Hall, radiative and Auger recombination.
The steady-state Shockley-Read-Hall recombination rate density is given by:
\be
   R_{\rm SRH} = \frac{np - n_i^2}{\tau_p(n+n_1) + \tau_n(p+p_1)} \label{eq:SRH}
\ee
where $n_i$ is the material intrinsic carrier density, given by $n_i=\sqrt{N_C N_V} \exp\left(-E_g/(2k_BT)\right)$.
The equilibrium Fermi energy at which $n=p=n_i$ is the instrinsic energy level $E_i$.  We specify the defect energy level $E_T$ relative to $E_i$ (see Fig. \ref{fig:energy}),
so that the expressions for $n_1$ and $p_1$ in Eq. \ref{eq:SRH} are given by:
\begin{align}
n_1 &= n_i \exp\left(\frac{ E_T}{k_BT}\right) ,\\
p_1 &= n_i \exp\left(-\frac{ E_T}{k_BT}\right)
\end{align}
$\tau_{n,(p)}$ is the bulk lifetime for
electrons, holes. It is given by
\be
   \tau_{n,p} = \frac{1}{N_T v^{\rm th}_{n,p} \sigma_{n,p}}
   \label{tau}
\ee
where $N_T$ is the three-dimensional trap density, $v^{\rm
th}_{n,p}$ is the thermal velocity of carriers ($v^{\rm th}_{n,p} = \sqrt{3k_BT
/m_{n,p}}$ with $m_{n,p}$ the electron/hole effective mass), and $\sigma_{n,p}$
is the capture cross-section for electrons, holes.

\begin{figure}
\includegraphics[width=0.35\textwidth]{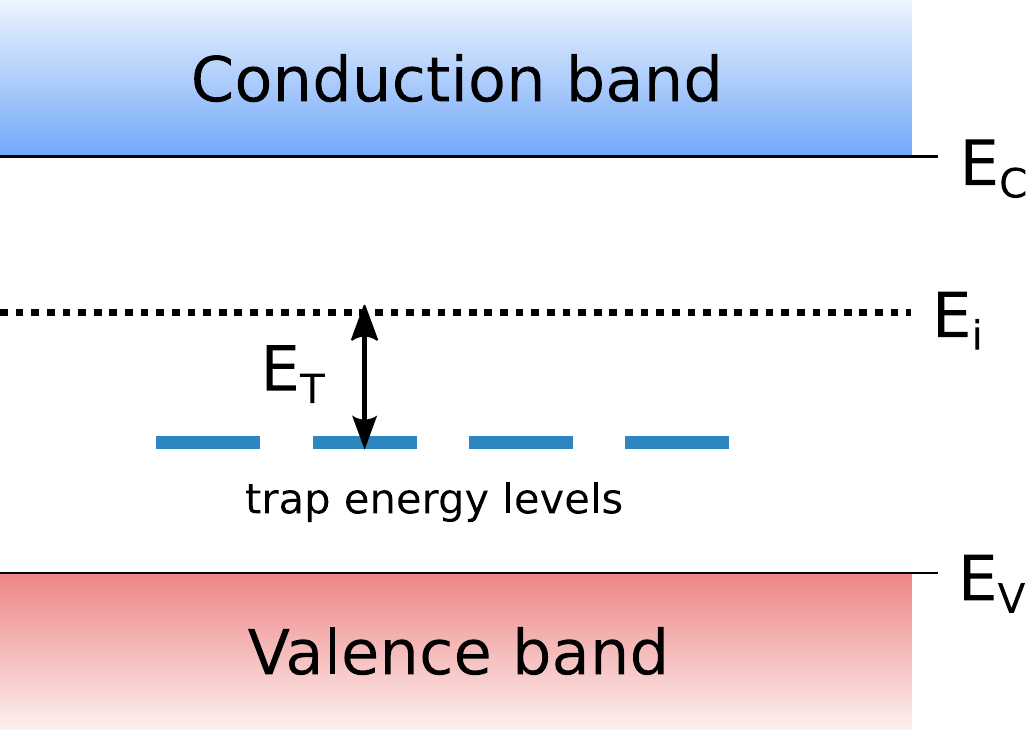}
\caption{\label{fig:energy} Depiction of energy levels of defect states.  The intrinsic energy level is $E_i = E_g/2 - k_BT/q \ln\left(N_C/N_V\right)$, as measured from the valence band edge. }
\end{figure}

The radiative recombination has the form
\be
   R_{\rm rad} = B (np - n_i^2)
\ee
where $B$ is the radiative recombination coefficient of the material. The
Auger mechanism has the form
\be
   R_{\rm A} = (C_n n + C_p p) (np - n_i^2)
\ee
where $C_n$ ($C_p$) is the electron (hole) Auger coefficient.

\bigskip
\subsubsection{Planar defects}
Sesame has been created with the intent of studying extended defects
in solar cells, such as grain boundaries and sample surfaces.
These extended planar defects are represented by a point in a 1-d model,
a line in a 2-d model, and a plane in a 3-d model.  The extended defect energy level
spectrum can be discrete or continuous.
For a discrete spectrum, we label the defect with the subscript {\it d}.
The occupancy of the defect level $f_d$ is given by~\cite{PhysRev}
\be
    f_d = \frac{S_n n + S_p p_d}{S_n(n+ n_d) + S_p(p+ p_d)}
    \label{fd}
\ee
where $n$ ($p$) is the electron (hole) density at the spatial location of the
defect, $S_n$, $S_p$ are recombination velocity parameters for electrons
and holes respectively. $n_d$ and $p_d$ are
\begin{align}
    n_d &= n_i \exp\left(\frac{E_d}{k_BT}\right)\\
    p_d &= n_i \exp\left(-\frac{E_d}{k_BT}\right)
\end{align}
where $E_d$ is calculated from the intrinsic level $E_i$.

The electron/hole recombination velocity are related to the electron/hole capture cross section and the defect density $\rho_d$ according to:
\be
S_{n,p}= \rho_d \sigma_{n,p} v^{\rm th}_{n,p}.
\ee

The defect recombination is of Shockley-Read-Hall form:
\be
   R_d = \frac{S_nS_p(np - n_i^2)} {S_n(n + n_d) + S_p(p + p_d)}.
\ee

The charge density given of a single defect depends on the defect type (acceptor vs. donor)
\be
    Q = q\rho_d\times \begin{cases} (1-f_d)~~~~~~~{\rm donor}\\ (-f_d)~~~~~~~~{\rm acceptor} \end{cases}
\ee
where $\rho_d$ is the defect density of state at energy $E_d$.
Multiple defects are described by summing over defect label $d$, or performing an integral over
a continuous defect spectrum.

\subsection{Boundary conditions at the contacts}

For a given system definition, Sesame first solves the equilibrium problem.  In equilibrium, the quasi-Fermi level
of electrons and holes levels are equal and spatially constant.  We choose an energy reference such that
in equilibrium, $E_{F_p}=E_{F_n}=0$.  The equilibrium problem is therefore reduced to a single variable $\phi^{\rm eq}\left({\bf r}\right)$.
Sesame employs both Dirichlet and von Neumann equilibrium boundary conditions for $\phi^{\rm eq}$, which we discuss next.

\subsubsection{System in thermal equilibrium}

Sesame uses Dirichlet boundary conditions as the default.  This is the appropriate choice apply when the equilibrium charge density at the contacts is known {\it a priori}.  This applies for Ohmic and ideal Schottky contacts.  For Ohmic boundary conditions,
the carrier density is assumed to be equal and opposite to the ionized dopant density at the contact.
For an $n$-type contact with $N_D$ ionized donors at the $x=0$ contact ({\it i.e.} no free excess carriers at the contact), Eq. \ref{n} yields the expression for $\phi^{\rm eq}(x=0)$:
\be
q\phi^{\rm eq}\left(0,y,z\right) = k_BT\ln\left(\frac{N_D}{N_C}\right) - \chi\left(0,y,z\right)
\ee
Similar reasoning yields expressions for $q\phi^{\rm eq}$ for $p$-type doping and at the $x=L$ contact.

For Schottky contacts, we assume that the Fermi level at the contact is equal to the Fermi
level of the metal.  This implies that the equilibrium electron density is $N_C\exp\left[-\left(\Phi_M-\chi\right)/k_BT\right]$,
where $\Phi_M$ is the work function of the metal contact.  Eq. \ref{n} then yields the expression for $\phi^{\rm eq}$
(shown here for the $x=0$ contact):
\be
q\phi^{\rm eq}\left(0,y,z\right) = -\Phi_M|_{x=0~{\rm contact}}
\ee
An identical expression applies for the $x=L$ contact.

Sesame also has an option for von Neumann boundary conditions, where it's assumed that the
electrostatic field at the contact vanishes:
\be
   \frac{\partial \phi^{\rm eq}}{\partial x}(0, y, z) = \frac{\partial \phi^{\rm eq}}{\partial
   x}(L, y, z) = 0.
   \label{bc1}
\ee
The equilibrium potential $\phi^{\rm eq}$ determines the equilibrium densities $n^{\rm eq},~p^{\rm eq}$ according
to Eqs. \ref{n} and \ref{p} with $E_{F_n}=E_{F_p}=0$.

\subsubsection{System out of thermal equilibrium}

Out of thermal equilibrium, Dirichlet boundary conditions are imposed on the
electrostatic potential. For example, in the presence of an applied bias
$V$ at $x=L$, the boundary conditions are
\begin{align}
    \label{bc2}
   \phi(0, y, z) &= \phi^{\rm eq}(0,y,z)\\
   \phi(L, y, z) &= \phi^{\rm eq}(L,y,z) + qV
    \label{bc3}
\end{align}
where $\phi^{\rm eq}$ is the equilibrium electrostatic potential.

\bigskip
For the drift-diffusion equations, the boundary conditions for carriers at
charge-collecting contacts are parameterized with the
surface recombination velocities for electrons and holes at the contacts,
denoted respectively by $S_{c_n}$ and $S_{c_p}$:
\begin{align}
   \label{BCj1}
   J^x_n(0,y,z) &= qS_{c_n}^0 (n(0,y,z) - n^{\rm eq}(0,y,z))\\
   \label{BCj2}
   J^x_p(0,y,z) &= -qS_{c_p}^0 (p(0,y,z) - p^{\rm eq}(0,y,z))\\
   \label{BCj3}
   J^x_n(L,y,z) &= -qS_{c_n}^L (n(L,y,z) - n^{\rm eq}(L,y,z))\\
   J^x_p(L,y,z) &= qS_{c_p}^L (p(L,y,z) - p^{\rm eq}(L,y,z))
   \label{BCj4}
\end{align}

\subsection{Numerical implementation}

In this section we review the set of equations solved by Sesame and provide some details of their implementation in the one-dimensional case.

\subsubsection{Scharfetter-Gummel scheme}
\begin{figure}
\includegraphics[width=0.3\textwidth]{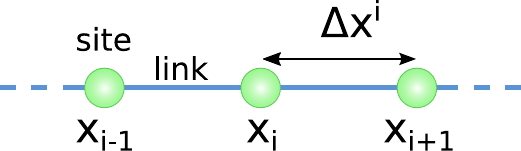}
\caption{\label{sites-links}Sites and links of the grid used in the
discretization of the drift diffusion and Poisson equations.}
\end{figure}

Sesame uses finite differences to solve the drift-diffusion-Poisson equations on a nonuniform grid.  Fig.~\ref{sites-links} shows our index-labeling convention for sites and links: link $i$ connects site $i$ and site $i+1$.  Site-defined quantities (such as density and electrostatic potential) are labeled
with a subscript denoting the site number. Link-defined quantities (such as electrical current and electric field) are labeled with a superscript denoting the link number.
\bigskip

We consider a one-dimensional system to illustrate the model discretization. First, we rewrite the current on link $i$ in semi-discretized form:

\begin{align}
    \label{jni}
    J_n^i & = q\mu_{n,i} n_i \left.\frac{\mathrm{d} E_{F_n}}{\mathrm{d}x}\right|_i \\
    J_p^i & = q\mu_{p,i} p_i \left.\frac{\mathrm{d} E_{F_p}}{\mathrm{d}x}\right|_i
   \label{jpi}
\end{align}

A key step to ensure numerical stability is to integrate Eqs.~(\ref{jni}) and (\ref{jpi}) in order to get a completely discretized version of the current $J^i_{n,p}$.  This discretization is known as the Scharfetter-Gummel scheme~\cite{Gummel}.  Here we give the final expressions for the hole current $J_p^i$ between sites $i$ and $i+1$:

\begin{align}
    J_p^i &=&  \frac{q}{\Delta x^i}
    \left(\frac{\psi_{p,i+1}-\psi_{p,i}}{\exp\left(\frac{\psi_{p,i+1}}{k_BT}\right)-\exp\left(\frac{\psi_{p,i}}{k_BT}\right)}  \right)
    \times \nonumber \\ && \mu_{p,i}
    \left[\exp\left(\frac{-E_{F_p,i+1}}{k_BT}\right)-\exp\left(\frac{-E_{F_p,i}}{k_BT}\right)\right].
   \label{Jpi}
\end{align}
where $\psi_p=q\phi + \chi + E_g - k_BT \ln(N_V)$ is the effective potential.  The electron current $J_p^i$ is given by:
\begin{align}
    J_n^i &=& -\frac{q}{\Delta x^i}
    \left(\frac{\psi_{n,i+1}-\psi_{n,i}}{\exp\left(\frac{-q\psi_{n,i+1}}{k_BT}\right)-\exp\left(\frac{-q\psi_{n,i}}{k_BT}\right)} \right)
    \times \nonumber \\ && \mu_{n,i}
    \left[\exp\left(\frac{E_{F_n,i+1}}{k_BT}\right)-\exp\left(\frac{E_{F_n,i}}{k_BT}\right)\right].
   \label{Jni}
\end{align}
where $\psi_n=q\phi+\chi+k_BT\ln(N_C)$.

In the limit where either $\delta\psi_{n(p)}\equiv -q\left(\psi_{n(p),i+1} - \psi_{n(p),i}\right)/k_BT$  or $\delta E_{F_{n(p)}}\equiv\left(E_{F_{n(p)},i+1}-E_{F_{n(p)},i+1}\right)/k_BT$ are smaller than $10^{-5}$ and $10^{-9}$, respectively, we replace the expressions for the current with a Taylor series expansion of the small parameter.  In the expansion, we evaluate the current up to second order in $\delta \psi_{n(p)}$, and up to first order in $\delta E_{F_{n(p)}}$.

Embedding a two-dimensional density into the three-dimensional model is formally accomplished with the use of a delta function. Numerically, the two-dimensional defect densities of states and the surface recombination velocities are divided by the size of the discretized grid at the position of the plane, and along the direction normal to the plane.

\begin{table}
\begin{tabular}{|l|c|c|}
  \hline
  Quantity & Expression & Value  \\ \hline
  Density & $N_0$ & $10^{27}~{\rm m^{-3}}$\\
  Mobility & $\mu_0$ & $10^{-4}~{\rm cm^2/\left(V\cdot s\right)}$\\
  Temperature & $T_0$  &  $300~{\rm K}$\\
  Energy & $k_BT_0$  &  $0.0258~{\rm eV}$\\
  Length      &    $\sqrt{\epsilon_0 k_B T/(q^2 N_0})$   &  $3.78\times 10^{-10}~{\rm m}$     \\
  Time      &    $\epsilon_0/(q \mu_0 N_0)$   &    $5.5\times10^{-14}~{\rm s}$   \\
  Gen. rate density      &    $N\mu_0 E_0/x_0^2$   &   $1.81\times10^{32}~{\rm 1/\left(m^2\cdot s\right)}$    \\
  Current      &    $\mu_0 N_0 k_B T/x_0$   &  $1.10\times 10^{10}~{\rm A/m^2}$     \\\hline
\end{tabular}
\caption{Quantities used to scale variables to dimensionless form. \label{scaling} }
\end{table}

\subsubsection{Newton-Raphson algorithm}
The discretization of Eqs.~(\ref{ddp1})-(\ref{ddp3}) leads to the system of
three equations for all sites of the discretized space (except boundary sites):
\begin{align}
    0 &= \frac{2}{\Delta x^i + \Delta x^{i-1}}\left(J_p^{i} -
    J_p^{i-1}\right) + G_i - R_i
    \\
    0 &= \frac{2}{\Delta x^i + \Delta
    x^{i-1}}\left(J_n^{i} - J_n^{i-1}\right) - G_i + R_i
    \\
    0  &=\rho_i +\frac{2}{\Delta x^i + \Delta x^{i-1}} \times \left[\left(\frac{\epsilon_{i+1} + \epsilon_{i}}{2}\right)\left(\frac{\phi_{i+1} - \phi_i}{\Delta x^i}\right) -\right. \nonumber \\ &\left.~~~~~~ \left(\frac{\epsilon_{i} + \epsilon_{i-1}}{2}\right)\left(\frac{\phi_{i} - \phi_{i-1}}{\Delta x^{i-1}}\right)\right]
\end{align}
Because we exchanged the carrier densities for the quasi-Fermi levels as the unknowns of the problem, we are therefore looking for the sets {$E_{F_n}, E_{F_p}, \phi$} at every grid point.

We use the Newton-Raphson method to solve the above set of equations: Given a general nonlinear function $f(x)$, we want to find its root $\bar x: f(\bar x)=0$.  Given an initial guess $x_1$, one can estimate the error $\delta x$ in this guess, assuming that the function varies linearly all the way to its root
\be
    \delta x= \left(\frac{\mathrm{d}f}{\mathrm{d}x} (x_1)\right)^{-1}f\left(x_1\right).
    \label{eq1d}
\ee
An updated guess is provided by $x_2 = x_1 - \delta x$.  The assumption of linear variation is key here, as if the guess $x_1$ is too far from the root, the convergence of the algorithm is very uncertain.

In multiple dimensions the derivative in \eq{eq1d} is replaced by the Jacobian.  In this case, \eq{eq1d} is a matrix equation of the form
\be
    \delta {\bf x} = A^{-1} {\bf F}\left({\bf x}\right)
    \label{eqNd}
\ee
where $\bf F$ is a vector function of the unknowns of the problem on all sites of the discretized space, and $A$ is the Jacobian matrix given by
\be
    A_{ij} = \frac{\partial F_i}{\partial x_j}.
    \label{Aij}
\ee
We find that convergence of the Newton-Raphson algorithm for this problem requires exact (analytically computed) values for the Jacobian.

\bigskip
In case the guess is far from the root we are looking for, the correction given by Eq. \ref{eq1d} can overshoot the solution.  A simple way to improve the convergence is to damp the corrections $\delta \bf x$ given by \eq{eqNd}. Inspired by an earlier work~\cite{Brown1976}, we found that the following procedure gives good results. For $\delta {\bf x_i} > 1$, we replace $\delta \bf x_i$ by
\be
    \delta {\bf \bar x_i} = \mathrm{sgn}(\delta {\bf
    x_i})\log\left(1+1.72|\delta {\bf x_i}|\right).
\ee
\end{appendices}

\end{document}